\documentclass[showpacs,preprintnumbers,twocolumn,amsmath,amssymb,superscriptaddress,natbib,prb]{revtex4}

\usepackage{graphicx}
\usepackage{dcolumn}
\usepackage{bm}

\begin{document}

\title{Anisotropic magnetic deflagration in single crystals of Gd$_5$Ge$_4$}

\author{S. V\'{e}lez}
    \email{svelez@ubxlab.com}
\author{J. M. Hernandez}
\author{A. Garc\'{i}a-Santiago}
\author{J. Tejada}

\affiliation{Grup de Magnetisme, Departament de F\'{i}sica Fonamental, Facultat de F\'{i}sica, Universitat de
Barcelona, c. Mart\'{i} i Franqu\`{e}s 1, planta 4, edifici nou, 08028 Barcelona, Spain}
\affiliation {Institut de Nanoci\`{e}ncia i Nanotecnologia IN$^2$UB, Universitat de Barcelona, c. Mart\'{i} i Franqu\`{e}s 1,
08028 Barcelona, Spain}

\author{V. K. Pecharsky}
\author{K. A. Gschneidner, Jr.}
\affiliation{The Ames Laboratory, U.S. Department of Energy, Iowa State University, Ames, Iowa 50011-3020, USA}
\affiliation{Department of Materials Science and Engineering, Iowa State University, Ames, Iowa 50011-2300, USA}

\author{D. L. Schlagel}
\author{T. A. Lograsso}
\affiliation{The Ames Laboratory, U.S. Department of Energy, Iowa State University, Ames, Iowa 50011-3020, USA}

\author{P. V. Santos}
\affiliation{Paul-Drude-Institut f\"{u}r Festk\"{o}rperelektronik, Hausvogteiplatz 5-7, 10117 Berlin, Germany}

\date{\today}

\begin{abstract}
Experimental evidence of the anisotropy of the magnetic deflagration associated with the low-temperature first order antiferromagnetic (AFM) $\rightarrow$ ferromagnetic (FM) phase-transition in single crystals of Gd$_5$Ge$_4$ is reported. The deflagrations have been induced by controlled pulses of surface acoustic waves (SAW) allowing us to explore both the magnetic field and temperature dependencies on the characteristic times of the phenomenon. The study was done using samples with different geometries and configurations between the SAW pulses and the direction of the applied magnetic field with respect to the three main crystallographic directions of the samples. The effect of temperature is nearly negligible, whereas observed strong magnetic field dependence correlates with the magnetic anisotropy of the sample. Finally, the role of the SAW pulses in both the ignition and formation of the deflagration front was also studied, and we show that the thermal diffusivity of Gd$_5$Ge$_4$ must be anisotropic, following $\kappa_a>\kappa_b>\kappa_c$.
\end{abstract}


\maketitle

\section{INTRODUCTION}
The Gd$_5$Ge$_4$ intermetallic compound, together with other Si doped Gd${_5}$(Si$_x$Ge$_{1-x}$)$_4$ alloys, has attracted considerable attention over the last few years, principally due to their unusual giant magnetocaloric properties \cite{GdSiGeGMC,PRLPech2000,PRLPech2003,Sharath}. This effect is associated with a first-order AFM$\leftrightarrow$FM phase transition that occurs simultaneously with a structural transformation \cite{Mor2000,Lev2001}. The rich phenomenology of these transitions in Gd$_5$Ge$_4$ has been broadly studied as a function of temperature and magnetic field using both polycrystalline and single crystalline samples \cite{Lev2002,Lev2004,Chatt,Tang,Ouyang,Felix2004,Ouyang2011}. At temperatures exceeding $T\sim$ 20 K, the isothermal field-driven AFM$\leftrightarrow$FM transitions can be continuously reproduced but, when the sample is cooled below $T\sim10$ K at zero magnetic field, the field-driven AFM$\rightarrow$FM transition becomes irreversible. Several reported magnetic behaviors, such as the glassy properties\cite{Glass,Dev}, have suggested that the magnetocrystallographic ground state of this system at zero magnetic field could be the FM O(I) [O(I) is a common notation of the low-volume orthorhombic polymorph of Gd$_5$Ge$_4$, see Ref. \onlinecite{Refa} for details], whereas the initial AFM O(II) state [the O(II) is a common notation of the high-volume orthorhombic polymorph of Gd$_5$Ge$_4$], which is stable in a zero field cooled (ZFC) process, is due to the kinetic arrest of the O(II) crystal structure. However, recent first principles modeling \cite{Paudyal2007,PaudyalJPCM,Paudyal2010} of the free energy of the different O(I) and O(II) magnetic phases, have pointed out that the ground state of this crystal must be the AFM O(II).

At low temperatures, the first order field-driven AFM$\rightarrow$FM phase transition can proceed in two different fashions. Usually, this transition is rather gradual and takes place over a wide range of magnetic fields, but when the sample is large enough or the magnetic field is swept at high rates, this phase transformation is abrupt, which can be identified in the $M-H$ cycle as a magnetic jump \cite{Lev2001}. Historically, such magnetic discontinuities have been called \emph{magnetic avalanches} and they have been also observed in other materials \cite{Paulsen1995,Barco1999,Mahendiran,Ghiv2004,Fisher,Rana,Hardy,Suresh} which also exhibit a giant magnetocaloric effect \cite{Ghiv2004,Fominaya} related to a transition from a kinetically-arrested state to the magnetic equilibrium \cite{Sharath,Ghiv2005}. The dynamics of the magnetization of the sample during such transitions have been reported first in molecular magnets \cite{YokoPRL,QMD,McHugh2007,AHPumping,McHugh2009a,McHugh2009b,Decelle,Macia2009,Vill}, and later in manganites\cite{MaciaMan,MaciaFinger,MaciaRTC} and polycrystalline samples of Gd$_5$Ge$_4$ \cite{velezGd}. For all of these materials, it was found that a phase-transition front is formed and propagates through the sample at a constant speed on the order of a few m/s. The strong similarities between this magnetic phenomenon and a chemical combustion \cite{Combustion} have lead the scientific community to call it \emph{magnetic deflagration}.

In magnetic deflagration, the role of fuel is played by the energy difference between the metastable and the stable states of the system, namely $\Delta E$. This energy difference is related to the intrinsic energy of the metastable ordered magnetic phase plus the Zeeman energy, that comes from the interaction of an external magnetic field $H$ with the spins in the system. On the other hand, the rate of heat transferred from the region of \emph{burning} spins to their \emph{flammable} neighbors is controlled by i) the energy barrier to be overcome by the metastable spins, $U$; ii) the so-called characteristic time attempt $\tau_0$; iii) the thermal diffusivity $\kappa$; and iv) the fraction of flammable spins $n_m$. The main difference between chemical combustion and magnetic deflagration is that for the latter the source of energy is the reordering of the spins of the system instead of an irreversible chemical reaction. Therefore, magnetic deflagration becomes of special interest due to the non-destructive character of the process that allows complex phenomena of deflagration to be explored under different conditions.

The theory of magnetic deflagration \cite{TheoryMD} determines the instability condition that leads a typical broad transition to the occurrence of a deflagration process. When the rate of the thermal jump over the energy barrier for a single metastable spin, $\Gamma$, exceeds some critical value, $\Gamma_c$, the nucleation of the deflagration front and the subsequent thermal runaway should take place. This critical rate can be written as
\begin{equation}\label{eq:gammac}
\Gamma_c = \frac{8 k(T) k_B T^2}{U(H) \Delta E(H)n_{m} l^2},
\end{equation}
where $l$ is some characteristic length, $k(T)=\kappa(T)C(T)$ is the thermal conductivity, and $C(T)$ is the specific heat. The front of propagation is identified as the \emph{flame} of the process, whose characteristic size is $\delta \sim \left[\kappa(T_f)/\Gamma(T_f)\right]^{1/2}$, where $T_f$ is the corresponding temperature of the flame, which is given by
\begin{equation}\label{eq:tf}
T_f =\frac{\Theta_D}{\pi}\left[\frac{5n_m\Delta E(H)}{3k_B\Theta_D}\right]^{1/4},
\end{equation}
where $\Theta_D$ is the Debye temperature. Finally, in the approximation of a planar burning-front, the speed of the flame is
\begin{equation}\label{eq:speed}
v(H) = \left[\frac{\kappa(T_f)}{\tau_0}\cdot{\frac {4k_BT_f(H)}{U(H)}}\right]^{1/2}\exp\left[\frac{-U(H)}{2 k_B T_f(H)}\right].
\end{equation}

Note that Eq. (1) can be used to establish the threshold condition for the occurrence of spontaneous magnetic deflagration in a field-sweep experiment \cite{YokoPRL,AHPumping,Macia2009,Decelle,MaciaMan,velezGd}. However, in such experiment, it becomes a difficult task to explore the field dependence of the speed of the flame, $v(H)$. To solve that, experimentalists have developed techniques to study the magnetic deflagration using controlled experimental conditions. These experiments consist, basically, of sending a heat pulse that acts as a spark of flame that ignites the deflagration. Attached resistors\cite{Vill,McHugh2007,McHugh2009a,McHugh2009b}, electrical contacts made on the sample\cite{velezGd} or surface acoustic waves (SAW)\cite{QMD,MaciaMan} are examples of sources that can be used for this purpose.

However, the test of magnetic deflagration has been limited to only a single law of propagation. In the case of molecular magnets, the speed of the deflagration front is determined by the value of the magnetic field applied along the easy magnetization axis, whereas the transverse field affects the threshold conditions [see for example ref. \onlinecite{Macia2009}] via their unusual quantum properties \cite{TejadaMQT,BookMQTMM}. In the case of manganites, as well as polycrystalline samples of Gd$_5$Ge$_4$, there is no influence, excluding geometrical effects, of the direction of the applied magnetic field on the properties of the deflagration process, whose observed characteristics are the result of averaging the properties along the principal crystallographic axes of the sample due to their random distribution. Therefore, the goal of this work is to investigate whether the magnetic deflagration in single crystals of Gd$_5$Ge$_4$ is anisotropic, and what is the role of each crystallographic axis in both the formation and propagation of the deflagration front.

\section{EXPERIMENTAL SET-UP}

A large (approximately 4 mm diameter, 40 mm long) single crystal of Gd$_5$Ge$_4$ was grown using the tri-arc technique \cite{Refb}. The Gd metal used to prepare the stoichiometric polycrystalline charge weighing ~20 g total was prepared by the Materials Preparation Center of the Ames Laboratory \cite{Refc}, and it was at least 99.99 wt.\% pure with respect to all other elements in the periodic table.  The Ge was purchased from Meldform Metals, and it was 99.999 wt.\% pure. The as-grown single crystal was oriented using backscatter Laue technique.  Two different single crystals of Gd$_5$Ge$_4$ used in this work were cut from a larger single crystal using spark erosion. Their dimensions were as follows: sample 1: $S_1 \approx  1.17\times2.45\times1.04$ mm$^3$ and sample 2: $S_2 \approx 2.40\times1.29\times1.07$ mm$^3$ along the crystallographic directions $a$, $b$, and $c$, respectively.

Figure 1 shows the schematic of the experimental set-up used in our measurements. The Gd$_5$Ge$_4$ samples were mounted using non-magnetic commercial silicon grease on a piezoelectric device [Fig. 1(a)] specially designed to send SAW pulses to the sample. The excitation spectrum of SAW modes with this system is basically determined by the resonances of the interdigital transducer (IDT), whose values are multiple harmonics of a fundamental frequency $f_0\approx 111$ MHz (see, for example, ref. \onlinecite{QMD}, \onlinecite{MaciaMan} and \onlinecite{IDT} for more details). The device is placed inside a commercial SQUID magnetometer [Fig. 1(b)] capable of measurements at temperatures down to 1.8 K in magnetic fields up to 5T. The microwaves, for the SAW generation, were transported from an external commercial Agilent signal generator to the IDT placed inside the cryostat by means of a coaxial wire that introduces attenuation smaller than 10 dB.

The SAW pulse induced in the IDT propagates along the lenght of the piezoelectric substrate made of LiNbO$_3$. The crystallinity of the substrate, together with the geometry of the device, provides an amplitude profile of the SAW oscillations in the $XZ$ plane, being therefore the direction of oscillation of the SAW waves out of plane, that is parallel to the $y$ direction [see Fig. 1(a) for the definition of the axes]. Since $z$ is the largest side of the LiNbO$_3$ crystal and $x$ is the perpendicular direction, the amplitude profile of the SAW can be considered practically independent of $z$, but it depends on $x$ having a maximum at $x=0$ (that is, the position of the center of the IDT; see for example ref. \onlinecite{Profile}, and references therein, for more details). When desirable experimental conditions (a specific combination of $T$ and $H$) are reached and stable, a controlled SAW pulse is delivered to ignite the magnetic deflagration process in the Gd$_5$Ge$_4$ sample.

\begin{figure}[htbp!]
\includegraphics[scale=0.36]{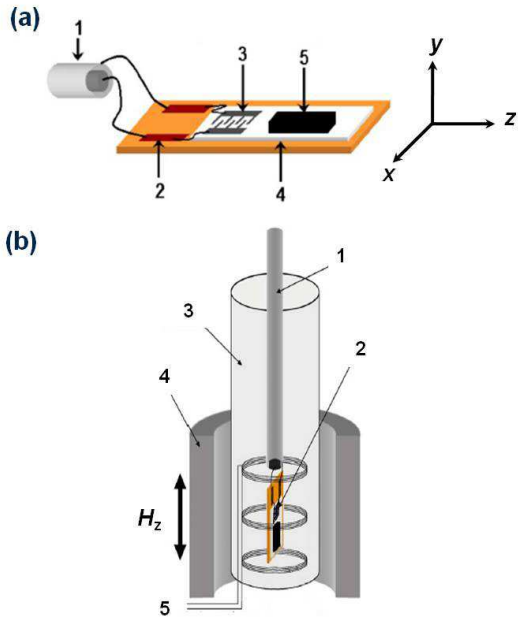}
\caption{Experimental set-up. (a) Schematic of the piezoeletric device used. 1, coaxial cable; 2, conducting stripes; 3, IDT; 4, LiNbO$_3$; 5, single crystalline sample of Gd$_5$Ge$_4$. (b) View of the spatial distribution of the piezoelectric set-up inside the SQUID magnetometer. 1, coaxial cable; 2, piezoelectric set-up; 3, sample holder; 4, superconducting coils for the field generation; 5, magnetometer's pick-up coils.}
\end{figure}

The magnetic field was always applied along the $z$ direction of the sample holder [as shown in Fig. 1(b)], and the piezoelectric set-up was placed between the pick-up coils of the SQUID magnetometer. Two different techniques have been used to obtain magnetic measurements with this system. The first, and the typical mode, consists of taking dc magnetic measurements by moving the sample through the four pick-up coils of the magnetometer. It allows one to obtain the absolute magnetic moment of a given sample with a very high precision. This technique has been used i) to characterize the magnetic properties of the sample and ii) to verify the magnetic state of the sample before and after each induced deflagration. The limitation of this mode is the time it takes to perform each measurement, which is approximately $\sim30$ seconds. The second method consists of directly measuring the voltage from the SQUID-voltmeter without requiring any motion of the sample. Placing the center of the sample at the position of the inner coils [as shown in Fig. 1(b)], where the sensitivity of the system has its maximum, the voltage drop recorded is directly related to the magnetic state of the sample [i.e., $\Delta V \propto \Delta M$]. This technique allows us to monitor fast magnetic changes with a time-resolution better than 0.01 ms, so that we can measure the time evolution of the magnetization of the sample during a magnetic deflagration process.

Each single crystal has been studied under different sample/set-up configurations to elucidate the role of each crystallographic axis on the properties of the deflagration phenomenon. Due to geometrical restrictions, four different configurations were available to be explored for each sample. When the magnetic field is applied along the longest side of the crystal, the SAW pulse can be applied to either of the two shorter sides. On the other hand, if the applied field is along one of the shorter directions, the SAW pulse can be applied parallel to the other short side. We will refer to each sample configuration using the following notation: $S_i(x,y,z)$, where $i$ denotes the sample number, and $x$, $y$ and $z$, correspond to the orientation of each crystallographic axis of the sample with respect to the coordinate system of the sample holder shown in Fig. 1(a). For example, $S_2(a,b,c)$ refers to sample 2 with its $a$-axis parallel to the $x$-direction of the sample holder, $b$-axis parallel to the $y$-direction, and $c$-axis parallel to the $z$-direction, which is the direction of the magnetic field vector.

\section{RESULTS}

DC magnetic measurements of the field-driven AFM$\rightarrow$FM transition at low temperatures were carried out for both Gd$_5$Ge$_4$ samples and for each configuration after the sample was ZFC from $T=50$ K. Figure 2 shows the AFM$\rightarrow$FM transformation and the subsequent removal of the magnetic field in the FM state obtained at $T=2$ K for each independent crystallographic axis of the system using sample 1. The same results were obtained for sample 2 but these are not shown here for brevity. As the magnetic field is increased from zero, the linear slope of the $M(H)$ curves [note that the observed step around $H \sim 8$ kOe when the magnetic field is applied along the $c$-axis is associated with the spin-flop transition \cite{Lev2004}] suggest that the initial state of the sample is purely AFM, and it remains unchanged, until direction-specific critical magnetic field, $H_c$, is reached, whose values at $T=2$ K are 28 kOe, 23 kOe and 26 kOe for the $a$, $b$ and $c$ crystallographic axis, respectively. Above it, the AFM$\rightarrow$FM transformation is quite gradual, and it takes place over a field range $\Delta H\sim4$ kOe. The inset of Fig. 2 shows, specifically, the fraction $n_{AFM}(H)$ of metastable AFM spins around the transition. At higher temperatures, the critical field $H_c(T)$ decreases with the rate of $\text{d}H_c/\text{d}T\approx-1.5$ kOe/K in the range $2-8$ K. All of these results are in agreement with the previous data reported for single crystalline Gd$_5$Ge$_4$ samples\cite{Ouyang}.

\begin{figure}[htbp!]
\includegraphics[scale=1.05]{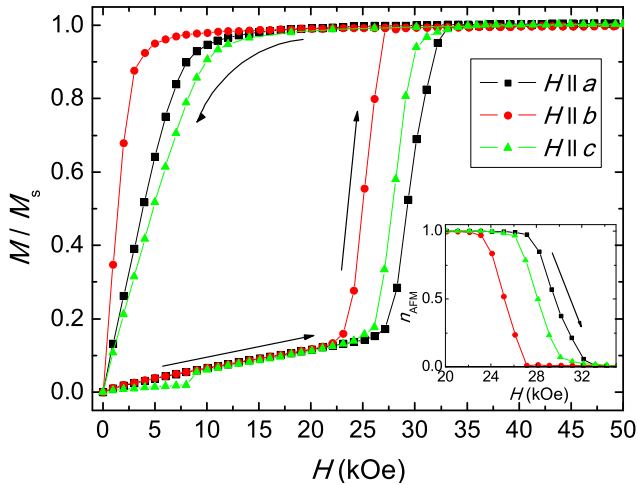}
\caption{(color online) Magnetic anisotropy of the field-driven AFM$\rightarrow$FM transition and the subsequent removal of the applied magnetic field measured at $T_0=2$ K after sample 1 was ZFC from $T=50$ K. The inset shows the fraction $n_{AFM}$ of the metastable AFM spins around the magneto-crystallographic transition.}
\end{figure}

After each crystallographic axis has been magnetically characterized, we proceeded to perform the deflagration experiments. Two control variables were i) the initial temperature $T_0$ and ii) the applied magnetic field for the ignition $H_{ig}$. For each deflagration measurement, the sample was ZFC from $T=50$ K to the desired temperature $T_0$. After that, the applied magnetic field was increased slowly to a selected ignition field $H_{ig}\lesssim H_c$, and then a SAW pulse of 100 ms width and 16 dB was delivered. As a consequence of the pulse, the sample is driven out of the initial equilibrium and for a certain range of experimental $T_0$ and $H_{ig}$ conditions, a magnetic deflagration was induced. The time evolution of the change in the magnetization of the sample, $\Delta M(t)$, was recorded from the SQUID-voltmeter, where $t=0$ corresponds to the delivery of SAW pulse.

Figure 3 shows the resulting $\Delta M(t)$ recorded for $S_1(c,a,b)$ configuration at $T_0=2$ K in the range of ignition fields at which the occurrence of magnetic deflagration was identified, and the initial value of $n_{AFM}$ was kept close to one. The data were normalized to the total magnetic drop observed for each case, $\Delta M_T$, which corresponds to the variation of $M$ when the spins of the sample change from the AFM state to the full FM state. This was confirmed from dc-magnetic measurements taken before and after each SAW pulse was delivered. For lower ignition fields [$H_{ig}<16.5$ kOe], the same kind of measurements performed revealed that no more than $\sim10\%$ of the spins of the sample become FM (not shown for simplicity). This abrupt change of the number of spins that transforms to the FM state with $H_{ig}$ is a typical feature that shows the self-maintenance of the deflagration process that utilizes the energy of the metastable spins for the occurrence of the full phase transformation when the threshold for the ignition of the deflagration is reached. In other words, in our experiment this threshold is reached for ignition fields from $16.5$ kOe to $22.0$ kOe, but at any smaller field, the effect of the SAW pulse is to induce transformation of only a fraction of spins into the stable FM state without deflagration. Moreover, the fraction of transformed spins decreases with decreasing $H_{ig}$.

\begin{figure}[htbp!]
\includegraphics[scale=0.43]{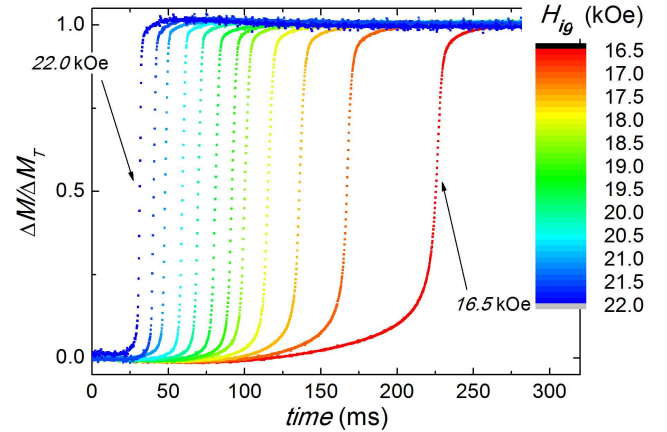}
\caption{(color online) Normalized time evolution of the change in the magnetization of the sample, $\Delta M(t)/\Delta M_T$, in the case of $S_1(c,a,b)$ for different ignition fields $H_{ig}$ at $T=2$ K at which a magnetic deflagration was observed. From right (red) to left (blue), the signals shown correspond to the data obtained for the values explored from $H_{ig}=16.5$ kOe to $H_{ig}=22.0$ kOe, respectively, in steps of 0.5 kOe. For each measurement, the sample was first ZFC from $T=50$ K, and then a $H_{ig}$ was applied. After that, a SAW pulse of 100 ms width was delivered at $t=0$.}
\end{figure}

\begin{figure*}[htbp!]
\center
\includegraphics[scale=2.05]{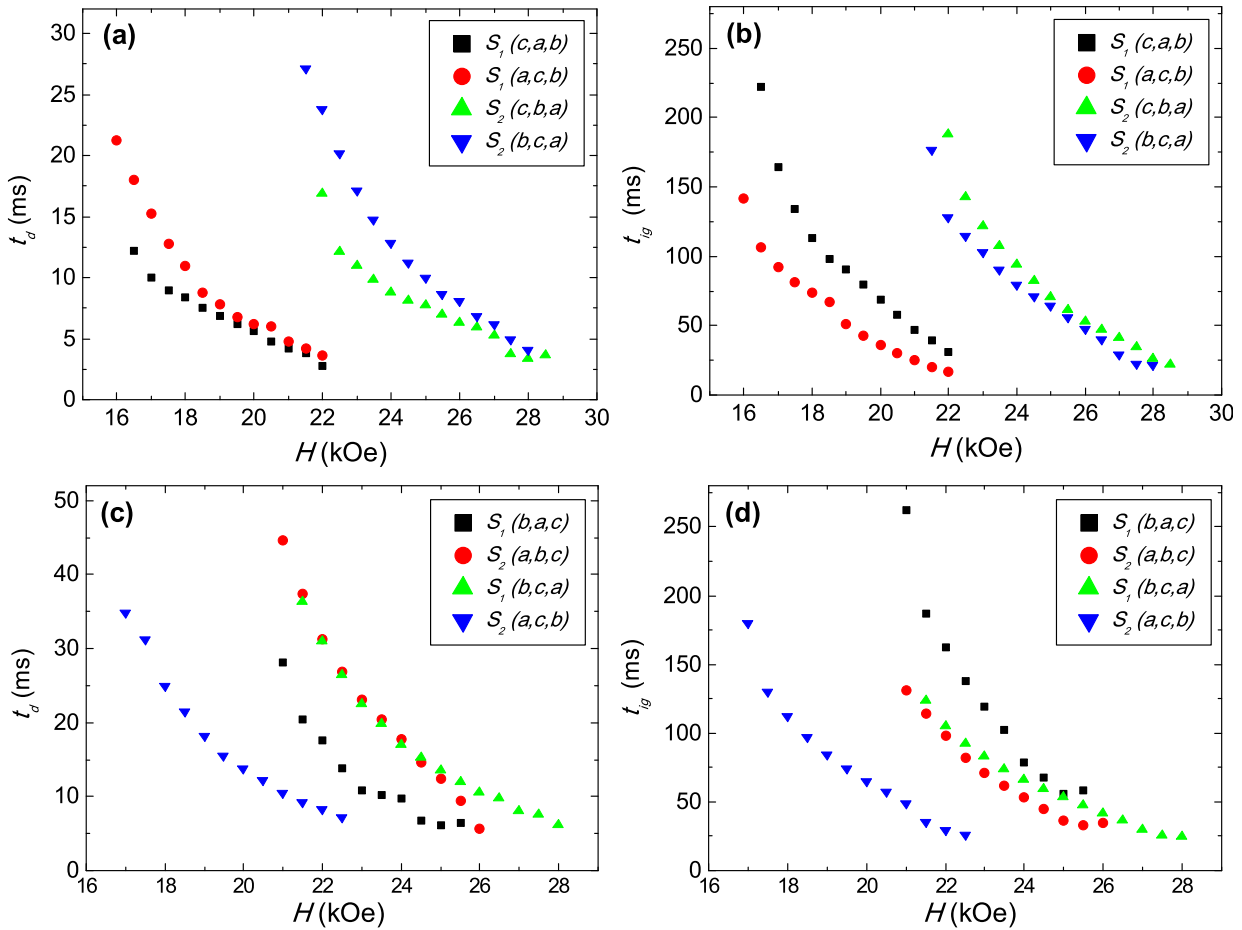}
\caption{(color online) The times of deflagration, $t_d$, and the ignition times, $t_{ig}$, obtained at different $H_{ig}$ for the two samples studied under different sample/set-up configurations split in two groups. (a) and (b) corresponds to $t_d$ and $t_{ig}$, respectively, when the applied field is along the longest side of the sample. (c) and (d) corresponds to the same data taken but for the perpendicular configurations.}
\end{figure*}

From the data shown in Fig. 3, two characteristic times related to the deflagration phenomenon can be identified for each curve: $t_{ig}$, that is the time that is required to reach the deflagration threshold after the SAW pulse has been switched on, and corresponds to the point at which the slope in the $\Delta M(t)$ curve exhibits an abrupt change; and $t_{d}$, defined as the subsequent time interval at which this fast change takes place due to the magnetic deflagration. Note that both characteristic times are strongly influenced by the ignition field, but no remarkable differences were observed when the experimental procedure was repeated at fixed $H_{ig}$ for a wide range of different $T_0$ values (not shown). It is an expected behavior that follows from the theoretical expressions given above. From Eq. 3, $v(H)$, which is related to $t_d(H)$, only depends on $H$ and is independent on $T_0$, on condition that $n_m$ is constant. This is our case because we restrict our experiments to $n_m=n_{AFM}\simeq1$. On the other hand, the value of $t_{ig}$ can be theoretically found solving the inequality $\Gamma(H,T)\geq\Gamma_c(H,T)$. Essentially, it is accomplished when $T$ is increased enough due to the delivery of a SAW pulse (in section IV.B we will show why the effect of the SAW pulse is to heat the sample), providing a certain $t_{ig}$ value for each experimental condition. However, testing this equation for different $T_0$ and $H_{ig}$ it can be shown that, whereas a small change in $H_{ig}$ provides a strong change in $t_{ig}$, the effect of $T_0$ is practically negligible for a wide range of experimental values. Additionally, the higher $T_0$, the smaller the capability of the system to retain the spins in the metaestable AFM state, reducing the experimental range over which the phenomenon of magnetic deflagration can be explored. Considering this, we have focused on the field dependencies of the properties of the deflagration phenomenon among the different crystallographic axes and sample configurations.

Figure 4 shows the characteristic times, $t_d(H_{ig})$ and $t_{ig}(H_{ig})$, obtained for the two samples and for the different configurations explored split in two groups according to different geometrical arrangements of the crystals in the sample holder: magnetic field applied along the longest side of the crystal [Fig. 4(a) and 4(b)] and magnetic field applied perpendicularly [Fig. 4(c) and 4(d)]. To avoid both the ignition and the extinction of the deflagration process that do not correspond well with the deflagration of a flame through the sample, we have used the middle part of the total magnetic change to identify $t_d$. In particular, we have taken the time elapsed between 25$\%$ and 75$\%$ of $\Delta M_T$ multiplied by 2 to rescale the data. Independently of the specific characteristic times observed for each configuration, several common features are worth noting. When $H_{ig}$ is close to $H_c$, related to the crystallographic axis of the sample that is parallel to the applied magnetic field, both the ignition and deflagration times are rather fast taking place only in a few tens of ms and few ms, respectively. However, as $H_{ig}$ moves down and away from $H_c$, both times increase progressively showing a non-linear dependence. Moreover, when $H_{ig}$ is reduced far from $H_c$, the energy supplied by the SAW pulse is no longer sufficient to reach the ignition threshold, and therefore, the deflagration does not take place. Moreover, depending on the configuration of the sample, a few deflagrations take place even after the SAW pulse has been switched off [for example, see $S_1(b,a,c)$ in Fig. 4(d) where five deflagrations exhibit $t_{ig}>100$ ms], indicating again the self-maintenance character of the deflagration phenomenon.

\section{Discussion}

There are two important features to take into account from the time values shown in Fig. 4. On one hand, for a certain ignition magnetic field $H_{ig}$, the observed values of both characteristic times of the magnetic deflagration mainly depend on the crystallographic direction of the sample along which the magnetic field is applied. For example, this follows from strong differences observed between the values obtained for $a$ and $b$ axes, where the field intervals at which the deflagrations take place do not overlap. However, notice that if the relative field value $H_{ig}-H_c$, where $H_c$ is the corresponding critical field along which the magnetic field is applied, is taken into account, the characteristic times observed become similar indicating that there is a correlation between the magnetic anisotropy of the sample and the observed deflagration times. On the other hand, the differences between the sets of data $S_1(a,c,b)$ and $S_1(c,a,b)$ [or $S_2(b,c,a)$ and $S_2(c,b,a)$], where the geometrical dispositions of the sample are equivalent (which would imply same lengths of propagation) and the applied magnetic field is along the same crystallographic axis, it suggests that additional effects on the properties of the magnetic deflagration may come from the crystallographic orientation of the sample in the $XY$ plane of the sample holder. Therefore, the aim of this section is to discuss these experimental observations within the framework of the theory of magnetic deflagration to show i) the connection between the magnetic anisotropy of the sample and the observed field dependencies and ii) the role of each crystallographic axis of the sample and the SAW pulse on the ignition and propagation of the flame.

\subsection{Comparison of the data with the Theory of Magnetic Deflagration}

For simplicity, and because $t_d(H)$ is a more reliable fingerprint than $t_{ig}$ of the deflagration phenomenon, we start our discussion focusing on these values. They must correspond to $t_d(H)\simeq l_p/v(H)$, where we have defined $l_p$ as the length along which the deflagration front propagates inside the sample and $v(H)$ is its speed. For a planar front propagating along one of the principal crystallographic axis of the sample, the time of propagation of the flame must follow:

\begin{equation}\label{eq:speed}
t_d^{2}(H) \approx {\frac{\tau_0}{4k_B}}\cdot\underbrace{\frac{l_p^2}{\kappa(T_f)}}_{g(l_p,\kappa)}\cdot\underbrace{{\frac {U(H)}{T_f(H)}}\exp\left[\frac{U(H)}{k_B T_f(H)}\right]}_{f(H)}.
\end{equation}

Omitting nonessential factors that cannot contribute to any observable difference of the values of $t_d(H)$ between the samples and configurations, all the experimental dependencies to be taken into account can be split in two different functions according to their origin: the term $f(H)$ is related to the magnetic field dependencies, i.e., the energy barrier $U(H)$ and the temperature of the flame $T_f(H)$ [see Eq. (2)], which essentially depends on $\Delta E(H)^{1/4}$ because in all of the data reported $n_{AFM} \simeq 1$. Notice that these field-dependent values correspond to the crystallographic axis along which the magnetic field is applied. On the other hand, geometrical contributions associated with the heat transport have been grouped in $g(l_p,\kappa)$: the propagation length $l_p$ and the thermal diffusivity $\kappa$, whose values correspond to the crystallographic direction along which the flame propagates. In the case that it does not propagate only along one direction, the $g(l_p,\kappa)$ function should be modified with an appropriate combination of the values of $\kappa$ and $l_p$ for the axes involved in the process (we will further discuss this question in section IV.B). The characteristic time attempt, $\tau_0$, has been considered as a constant because it is typically a global parameter that characterizes the time-flip of all the spins of a solid and it should be independent of the direction at which the magnetic field is applied.

For any set of data, the main contribution to $t_d(H)$ comes from the $f(H)$ function due to the strong field-dependent term $\exp[U(H)/k_BT_f(H)]$. Notice that this exponential dependence can explain the non-linearity observed in Fig. 4. Since $T_f(H) \propto \Delta E(H)^{1/4}$, the values of $t_d(H)$ should be basically determined by the field dependence of the energy barrier, $U(H)$, whereas $T_f(H)$ may be considered as a constant for each set of data due to the limited range of experimental fields explored. Additionally, due to the weak magnetic anisotropy of the material, the values of $T_f$ should be similar for all crystallographic directions in our experiments. In this context, both the scaling of $t_d$ through the relative value of $H_{ig}$ with respect to the corresponding $H_c$, and the magnetic anisotropy of the sample, can be directly explained through the anisotropy of the field dependence of the energy barrier, $U(H)$. To better understand it, for a given temperature and crystallographic direction, notice that $H_c$ correspond to the field at which the effective energy barrier $U(H)$, is reduced enough to get the characteristic time attempt $\tau=\tau_0\exp[U(H)/k_BT]$ of the order of the time of measurement \cite{BookMQTMM}. According to this, and taking into account that the field dependence of $U$ should be the same for all crystallographic directions, for same relative field values, $H_{ig}-H_c$, the strength of the energy barriers would be equal for the three axes, and therefore, similar values of $t_d$ are expected despite the differences that could arise from the geometrical term.

To be more precise in our further discussion on the geometrical contribution, $g(l_p,\kappa)$, to the deflagration values, we proceed to compare the data obtained for different configurations when the applied magnetic field is along the same crystallographic direction. In this case, the sets of data are controlled by the same $f(H)$ function. Additionally, for a similar geometrical dispositions of the sample in the sample holder, $l_p$ is expected to be similar. Therefore, the observed differences [for example, see Fig. 4(a)] must be attributed to an anisotropy of $\kappa$. However, to better understand this fact, we have to determine, first, what is the role of both the SAW pulse and sample configuration on the nucleation of the deflagration front and, second, through which length(s)/direction(s) the deflagration front propagates. It could also give a reason for the different observed $t_{ig}(H_{ig})$ values. However, before that, it is interesting to know whether the predictions of either the speed of the deflagration front or the values of the ignition time that arise from the theory of magnetic deflagration \cite{TheoryMD} are in the range of the observed experimental values.

From previously reported data \cite{velezGd} of the magnetic deflagration in a polycrystalline sample of Gd$_5$Ge$_4$, $T_f$ was estimated to be around 30 K. Taking into account the discussion given above about the $T_f(H)$ values of the different crystallographic axis, and that the magnetic properties of a polycrystal are averaged, we can assume that $T_f\sim30$ K is a good estimate of the temperature of the flame in all observed deflagrations. Thus, the thermal diffusivity of the sample is estimated to be\cite{Fujieda} $\kappa(T_f)\sim 3\cdot 10^{-5}$ m$^2$/s. Taking $l_p \sim 1$ mm, $\tau_0\sim10^{-7}$ s, and that in our experimental field range $U(H)$ should be around $200-300$ K, one gets from Eq. (4), $t_d \sim 10$ ms, which is in a good agreement with our observations. The other interesting parameter to test is the width of the flame $\delta$, which can be found through $\delta\sim\left[\kappa(T_f)/\Gamma(T_f)\right]^{1/2}$. Here, $\Gamma(T_f)=\tau_0^{-1}\exp\left[-U/k_BT_f\right]$, so the upper limit for the width of the flame is estimated to be $\delta_{max}\sim0.1$ mm, which means that such a flame can be formed and can propagate inside the sample.

For the ignition time values, they can be estimated solving the inequality $\Gamma\gtrsim\Gamma_c$ [Eq. (1)]. From Ref. \onlinecite{Fujieda}, $k(T)\sim8$ J/s$\cdot$K. Using all of the estimated values given above, the condition $\Gamma\sim\Gamma_c$ should be accomplished for $T_{ig}\sim 12$ K. In other words, the deflagration process may take place if the temperature in some part of the sample is quickly raised above this threshold temperature \cite{TheoryMD}. Although in our experiments this cannot be done fast enough, we may assume that this condition could be accomplished if a certain volume of spins of the sample, of the order of the spins contained in the burning front, are heated around this temperature. Since the upper limit for the width of the deflagration front is $\delta_{max}\sim0.1$ mm, the upper limit for this volume is on the order of 10\% of the total volume of the sample. Then, we proceed to estimate the thermal rise experimented by this volume due to a SAW pulse of $\Delta t = 100$ ms. Assuming a good transfer between the SAW pulse and the sample, the transferred energy is estimated to be $E_{SAW}=P\cdot \delta t\sim10$ mW $\cdot$ 0.1 s = 1 mJ. On the other hand, the heating of $N_{\delta}$ mols can be expressed as $dT=dE_{SAW}/C(T)\cdot N_{\delta}$. In our case, the mass to be heated is $m_{\delta}\sim1$ mg, so that $N_{\delta}\sim 10^{-6}$ mols. In the region of interest, $C(T)=\alpha T^3$, where $\alpha$ is a well-known constant whose value is\cite{Lev2001} 0.7 J/mol$\cdot$K$^4$. Therefore, the final temperature of this volume is given by $T_{\delta}^4=4\Delta E_{SAW}/\alpha m_{\delta}+T_0^4$, where $T_0$ is the initial temperature. Replacing all parameters in this equation with the known values, one gets $T_{\delta}\sim9$ K, which is in agreement with the expected ignition temperature, $T_{ig}$.

\subsection{Formation of the deflagration front}

Due to the location of the sample with respect to the piezoelectric device, and taking into account both the geometry of the sample and the quasi independence of the amplitude of the SAW oscillations in the $z$ direction, the spatial description of any deflagration phenomenon should be described in two dimensions, in which both the ignition and the propagation of the deflagration can be described in the $XY$ plane. Taking into account the profile of the SAW oscillations in the $x$ direction of the piezoelectric device, it is reasonable to assume that the energy is supplied to the sample mainly at the center of the bottom surface, defined as the (0,0) point in the $XY$ plane. On the other hand, at the range of our working frequencies, the phonon thermalization process should occur in less than 1 ms, whereas the characteristic times involved in our experiments are at least of the order of a few ms. All these features suggest that the interaction of the SAW pulses can be approximated as a spark of fire that essentially heats the sample at the (0,0) point.

The heat supplied to the sample, during a SAW pulse, diffuses in the $XY$ plane resulting in a thermal rise that depends on the position and the time elapsed from the ignition of the pulse. Essentially, each isotherm follows an ellipsoidal shape in the $XY$ plane, whose characteristic lengths should follow the relation $l_x/l_y\approx (\kappa_x/\kappa_y)^{1/2}$, where $l_i$ denotes the length of the $i$ axis of the sample referred to the (0,0) point and $\kappa_i$ is the thermal diffusivity of this axis. In the case of isotropic diffusion, the characteristic lengths $l_x$ and $l_y$ should be equal. Therefore, any deflagration ignited in any orientation of the sample in the XY plane has to exhibit the same characteristic times. However, if one crystallographic orientation exhibits higher thermal diffusivity, the supplied heat penetrates easier in such direction breaking the symmetry of the plane and, consequently, the resulting deflagration properties.

To illustrate the effect of the anisotropy of the thermal diffusion in our experiments, let us show what should happen with the characteristic times of the magnetic deflagration in the case of a square geometry [$L_x=L_y$, which corresponds to the case of the data shown in Fig. 4(a) and Fig. 4(b)] but with a strong anisotropy of the thermal diffusion [for example, suppose $\kappa_1>>\kappa_2$]. It can be easily found that the difference between the ignition times, when $\kappa_1$ is oriented parallel to the $x$ direction in comparison with the resulting times when it is oriented along the $y$ direction, approaches $t_{ig}(\kappa_{1\|}y)/t_{ig}(\kappa_{1\|}x) \rightarrow 2$ because, whereas in the first case the phase front should be generated when $l_x \rightarrow L_x/2$, in the second one $l_y \rightarrow L_y$ due to the point at which the SAW pulse is transferred to the sample. Therefore, the distance to be covered by the deflagration front formed is different in each case, with the ratio between the deflagration times approaching $t_{d}(\kappa_{1\|}y)/t_d(\kappa_{1\|}x)\rightarrow 1/2$.

Focusing on the data shown in Fig. 4(a), when the $y$ direction is along the $c$ crystallographic axis [$S_1(a,c,b)$ and $S_2(b,c,a)$], $t_d$ is higher than when either $b$ or $a$ axis are aligned in this direction [$S_1(c,a,b)$ and $S_2(c,b,a)$]. From the phenomenological point of view, the shape of the $t_d$ curves are the same when the $y$ direction is parallel to $c$, and on the other hand, the curves at which $c$ is parallel to $x$ are also similar between them. Moreover, Fig. 4(b) shows that for these sets of data, higher $t_{ig}$ values are obtained when $c$ is parallel to $y$ direction. These findings, together with the expected thermal diffusivity dependencies previously discussed, are indicative of smaller thermal diffusivity along the $c$ axis compared to $b$ and $a$ axis. From similar arguments, the values obtained in other configurations [for example, the data corresponding to the sets $S_1(a,b,c)$ and $S_2(b,a,c)$], suggest that $\kappa_a$ should be higher than $\kappa_b$.

Finally, note that, for $H$ applied along the same crystallographic direction, the differences of $t_d(H)$ values observed between configurations cannot be explained by a unique scaling factor. It should be attributed to a different ratio of distances covered by the deflagration front depending on the ignition field explored. At magnetic fields close to $H_c$, the system can be driven out of equilibrium easier. In other words, the deflagration front boundary should form close to the (0,0) point, whereas for smaller fields, its formation is more complicated and it should occur deeper inside the sample where the ignition can take place \cite{velezGd,TheoryMD}. Therefore, when $H \rightarrow H_c$, the phase front propagates all over the $XY$ surface. However, when $H$ explored is far away from $H_c$, the anisotropy of $\kappa$ implies that the deflagration front formed should be different depending on the configuration studied as has been explained before, and then, different propagation lengths and deflagration times are expected between them. In conclusion, the different observed field dependencies should be attributed to the geometric function, $g(l_p,\kappa)$, as a consequence of non-trivial interplay between how the front is generated and how it diffuses in the $XY$ plane.

\section {CONCLUSIONS}

To summarize, magnetic deflagrations associated with the first order AFM$\rightarrow$FM magneto-crystallographic transformations in single crystals of Gd$_5$Ge$_4$ have been induced by controlled SAW pulses. The study has been done for different experimental conditions and configurations between the SAW pulses and the applied magnetic field with respect to the crystallographic axes of the samples. As expected, the dynamics of the process fits well within the framework of magnetic deflagration theory, but the comparison of the data obtained between different configurations have revealed anisotropic character of the process associated with both magnetic and thermal properties of each of the three crystallographic axes of the sample. The main effect comes from the field dependence, which is correlated with the magnetic anisotropy of Gd$_5$Ge$_4$ through the anisotropic character of the field dependence of the energy barrier, $U(H)$.

The data obtained suggest that the thermal diffusivity is anisotropic, following $\kappa_a>\kappa_b>\kappa_c$. It plays an important role in the front formation and the subsequent propagation inside the sample due to the fact that the anisotropy of the thermal diffusion can be interpreted as hard and/or easy axes for the occurrence of the phenomenon. Reported previously anisotropy of electrical resistivity and magnetoresistance\cite{AMR}, sound propagation and elastic properties \cite{Refd} in these and related alloys, makes the conclusion about anisotropy of the thermal conductivity reasonable. However, additional experiments should be performed to confirm and further explore this property.

The role of the SAW pulses in the ignition of the magnetic deflagration has been also highlighted. The directionality of the SAW pulse transferred to the sample and the characteristic times evolved in the deflagration process suggest that the SAW pulses act as an unidirectional heater leading to the deflagration process ocurring in the perpendicular cross section of the sample. However, we note that for systems in which $t_{ig}(H)$ or $t_d(H)$ are less, or at least, of the order of 1 ms, the phonon-spin interactions could play an important role in the properties of the magnetic deflagration \cite{SAWint}. Finally, while this work concentrates on the anisotropy of the dynamics of the deflagration phenomena, the authors want to remark that simultaneous to the magnetic deflagration process, a structural change takes place in the system. Further studies should elucidate the very interesting physics of what is happening, principally, inside the magneto-structural burning front.

\section {ACKNOWLEDGEMENTS}

S. V. acknowledges financial support from Ministerio de Ciencia e Innovaci\'{o}n de Espa\~{n}a. J. M. H. and A. G. -S. acknowledge support from Universitat de Barcelona. J. T. acknowledges financial support from ICREA Academia. Work at the University of Barcelona was financially supported by the Spanish Government project MAT2008-04535 and Catalan Government project 2009SGR1249. Work at the Ames Laboratory is supported by the U.S. Department of Energy, Office of Basic Energy Science, Division of Materials Science and Engineering under Contract No. DE-AC02-07CH11358 with Iowa State University.

\bibliography{GdGeProperties,Deflagrations}

\end{document}